# IMPACT OF 4IR TECHNOLOGY AND ITS IMPACT ON THE CURRENT DEPLOYMENT

Bandar Alsulaimani[1] and Amanul Islam[2]

[1]Department of Electrical Engineering, King Fahad University of Petroleum and Minerals (KFUPM), Saudi Arabia
[2]Department of Computer Science and Information Technology, University of Malaya, Malaysia

## ABSTRACT

*The Fourth Industrial Revolution represents a fundamental change in how we live, work, and relate to one another. It is a new chapter in human development with remarkable technological advancements comparable to those of the first, second, and third industrial revolutions. These developments are fusing the physical, digital, and biological worlds in ways that hold great promise as well as the possibility of great danger. The way that modern people live and work is changing as a result of disruptive technologies and trends including the Internet of Things (IoT), robotics, virtual reality (VR), and artificial intelligence (AI). This is known as the fourth industrial revolution. Industry 4.0 refers to the incorporation of these technologies into production processes. In this article, we discussed the history of 4IR technology, its impact of 4IR technology, and its impact on the current deployment.*

## 1. INTRODUCTION

Due to increased interconnection and smart automation, the Fourth Industrial Revolution, also known as Industry 4.0, envisions fast change in technology, industries, and societal patterns and processes in the twenty-first century[1]. The word has been widely used in scientific literature, and Klaus Schwab, the World Economic Forum's Founder and Executive Chairman, popularized it in 2015. According to Schwab, the changes are more than merely efficiency gains; they represent a significant transition in industrial capitalism[2].

The merging of technology such as artificial intelligence and gene editing with advanced robotics, which blurs the borders between the physical, digital, and biological worlds, is a part of this phase of industrial development[3].

Through ongoing automation of old manufacturing and industrial methods, the use of current smart technologies, large-scale machine-to-machine communication (M2M), and the internet of things(IoT), major alterations in how the global production and supply network runs are taking place[4]. Increased automation, improved communication and self-monitoring, and the usage of smart technologies that can evaluate and diagnose issues without the need for human interaction are all benefits of this integration[5].

      53

It also reflects a social, political, and economic transformation from the digital era of the late 1990s and early 2000s to an era of embedded connection marked by ubiquitous technological use (e.g., a metaverse) that alters how humans experience and understand the world around them. It asserts that, in comparison to humans' inherent senses and industrial abilities alone, we have built and are entering an enhanced social reality[6].

The purpose of this study is to discuss 4IR technology and how it affects current deployment. Sections 2 and 3 covered the background of 4IR technology as well as its effects and recent developments in computing technology. The effects of 4IR technology on business were briefly covered in Section 5, the effects of 4IR technology on the government were covered in Section 6, and the effects of 4IR technology on the people were covered in Section 7. The paper was ended in section 8.

## 2. HISTORY

A group of experts working on a high-tech strategy for the German government coined the term fourth industrial revolution. In a 2015 article published by Foreign Affairs, Klaus Schwab, executive chairman of the World Economic Forum (WEF), popularized the phrase to a wider audience[7]. At the 2016 World Economic Forum Annual Meeting in Davos-Klosters, Switzerland, the topic, was "Mastering the Fourth Industrial Revolution."

The Forum announced the opening of its Centre for the Fourth Industrial Revolution in San Francisco on October 10, 2016. Schwab's 2016 book was likewise about this, with the same title.This fourth era, according to Schwab, involves technologies that merge hardware, software, and biology (cyber-physical systems), with an emphasis on communication and connectivity [8]. Breakthroughs in emerging technologies such as robotics, artificial intelligence, nanotechnology, quantum computing, biotechnology, the internet of things, the industrial internet of things, decentralized consensus, fifth-generation wireless technologies, 3D printing, and fully autonomous vehicles, according to Schwab, will define this era[9].

**First Industrial Revolution**

Through the utilization of steam power and water power, manual production techniques were replaced by machines during the First Industrial Revolution. This refers to the period between 1760 and 1820, or 1840 in Europe and the United States because the adoption of new technology took a long time. Although it also had societal implications, such as a growing middle class, it had an impact on the iron industry, agriculture, and mining, as well as textile production, which was the first to embrace such innovations. At the time, it also had an impact on British industry[39].

**Second Industrial Revolution**

The period between 1871 and 1914, commonly referred to as the Technological Revolution, was marked by the establishment of massive railroad and telegraph networks that facilitated the speedier exchange of people, ideas, and electricity. The contemporary production line was developed in factories thanks to increased electricity. It was a time of rapid economic expansion and productivity increases, together with a rise in unemployment as many factory workers were replaced by machines [40].

**Third Industrial Revolution**

After the end of the two world wars, there was a pause in industrialization and technological development compared to earlier eras, which led to the Third Industrial Revolution, also known as the Digital Revolution.A decade later, the creation of the Z1 computer, which made use of binary floating-point numbers and Boolean logic, marked the start of increasingly sophisticated digital innovations.The supercomputer was the next important advancement in communication technologies, and with its widespread usage in the manufacturing process, equipment started to replace the need for human labor[41].

## 3. IMPACT OF 4IR TECHNOLOGY(NEW COMPUTING TECHNOLOGY)

Because of growing interconnectivity and smart automation, the Fourth Industrial Revolution, 4IR, or Industry 4.0, intellectualizes quick modification to technology, industries, and societal patterns and processes in the 21st century. Some 4IR new computing technology is smart factories,3D printing, smart sensors, etc[10]. A short discussion of smart factories,3D printing, and smart sensors is given below.

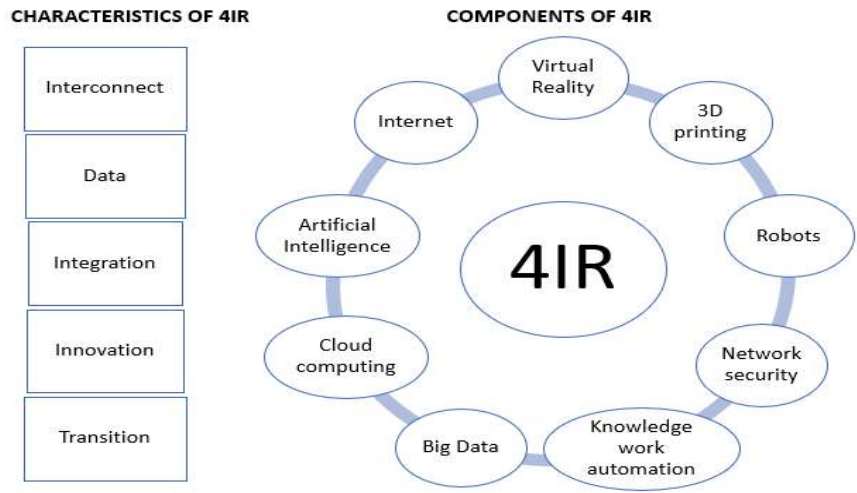

Fig-1. Components of 4IR technology

### 3.1. Smart Factory

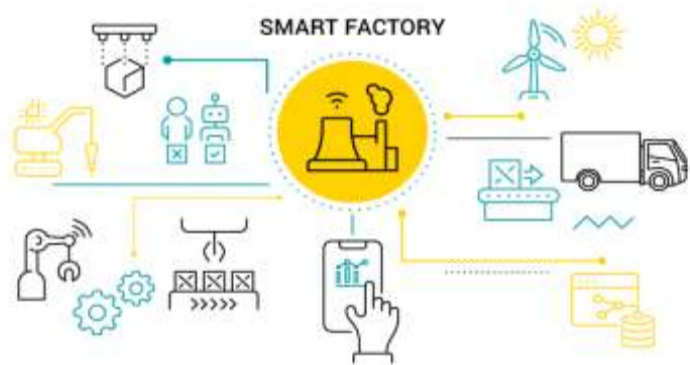

Fig-2. Example of a smart factory

The Smart Factory is a vision of a manufacturing environment in which production and logistical systems are arranged without the need for human interaction[11]. The Smart Factory isn't just a concept anymore. While various model factories illustrate what is possible, numerous businesses have previously explained how the Smart Factory works with examples[12].

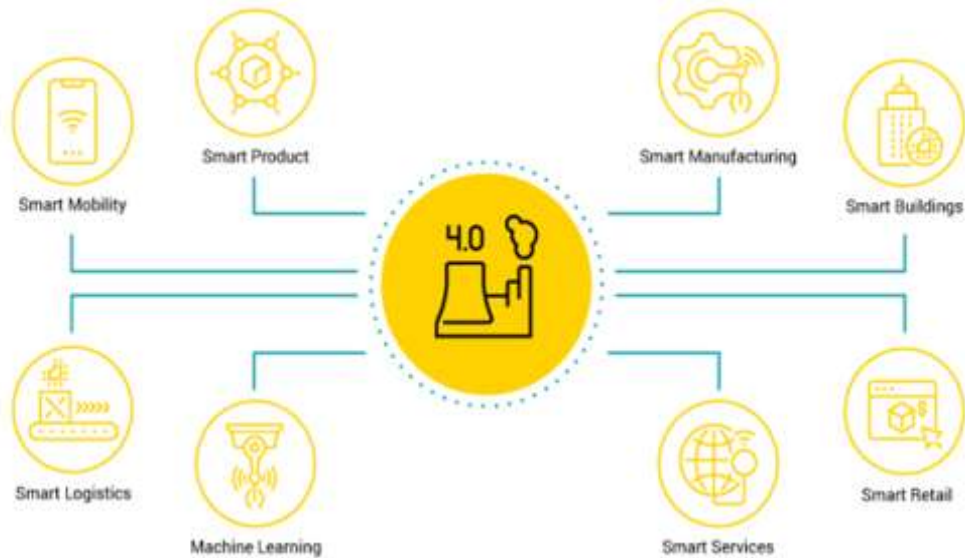

Fig-3. Technology of smart factory

Cyber-physical systems that connect with each Internet of Things and Services are the technical underpinnings on which the Smart Factory - the intelligent factory - is built. The data transfer between the product and the production line is a crucial step in this process. This makes it possible for the Supply Chain to be connected much more effectively and for any manufacturing environment to be better organized [13].

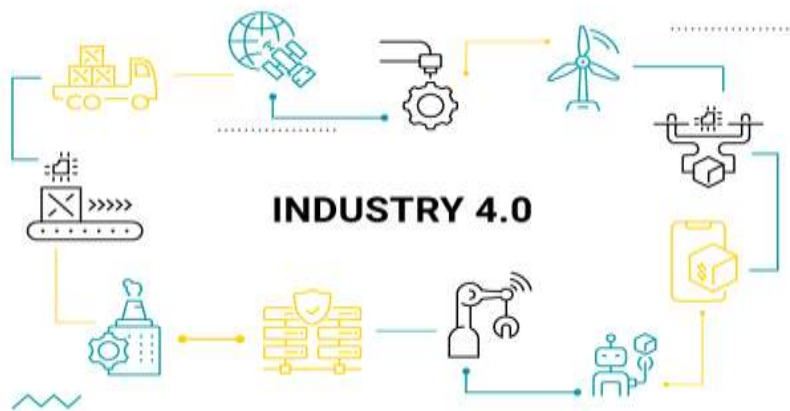

Fig-4. 4IR technology in smart factory

The Fourth Industrial Revolution fosters what has been called a "smart factory".Cyber-physical systems control physical processes, simulate the real world, and make decentralized choices within modularly built smart factories. Cyber-physical systems communicate and work together with humans and other cyber-physical systems across the internet of things in real-time, both within and between organizational services provided and used by value chain players [14].

## 3.2. 3D Printing

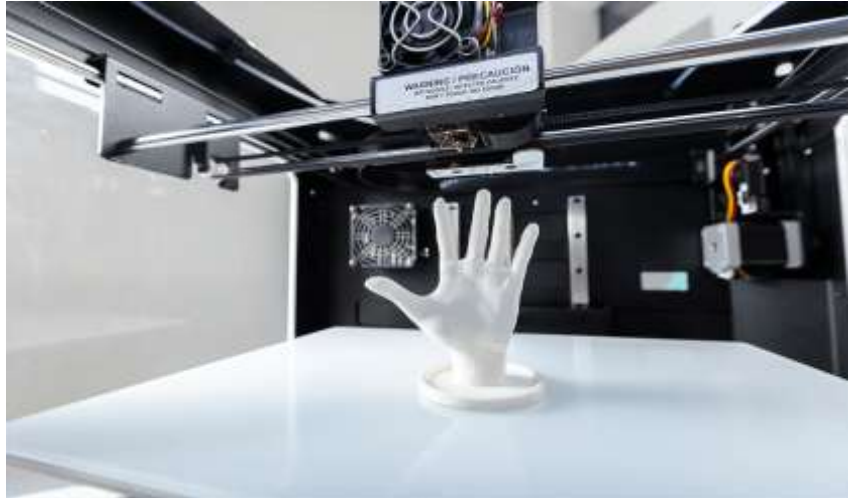

Fig-5. Example of 3d printing

It is believed that 3D printing technology will play a significant role in the Fourth Industrial Revolution. The ability to manufacture numerous geometric forms and the simplification of the product design process are two benefits of 3D printing for the industry. Additionally, it is comparatively eco-friendly. It can help cut down on lead times and overall production costs in low-volume production[15]. Furthermore, it can improve flexibility, lower storage expenses, and aid the organization in adopting a mass customization business plan. Moreover, printing spare parts in 3D and installing them locally can reduce reliance on suppliers and shorten the lead time for supplies[16].

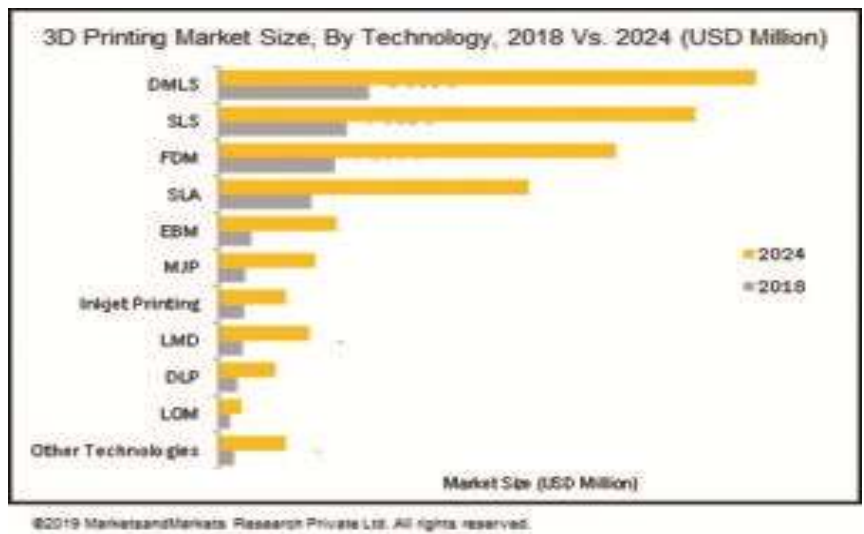

Fig-6. Market trends of 3d printing

There are several restrictions on the projected expansion of the 3D printing industry. First, even if the cost of equipment has decreased dramatically since the technology's inception, there are still other considerations, such as energy expenditures. A global survey found that "there are many elements that contribute to the greater cost of the 3D printing devices[17]. The energy required by 3D printing to create products is significant, to name a few. For instance, compared to injection molding equipment, the energy released by various 3D printing methods might use up to 50 to 100 times more electricity[18].

The rate of change is the deciding element. One can identify a qualitative shift in the pace of development that ushers in a new historical period by relating the rate of technological advancement to socioeconomic and infrastructure changes that follow[19].

### 3.3. Smart Sensors

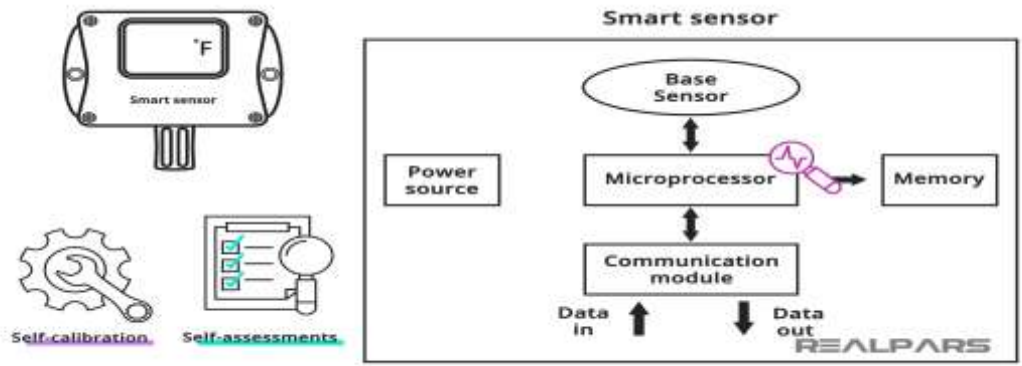

Fig-7. Example of a smart sensor

For Industry 4.0 and other "smart" megatrends including smart manufacturing, smart mobility, smart homes, smart cities, and smart factories, sensors and instrumentation are the driving forces of innovation[20].

Smart sensors are gadgets that provide data and enable additional functionality, such as self-auditing and self-configuration as well as condition monitoring of intricate operations. They greatly simplify installation work and aid in realizing a dense array of sensors because they have wireless communication capabilities[21].

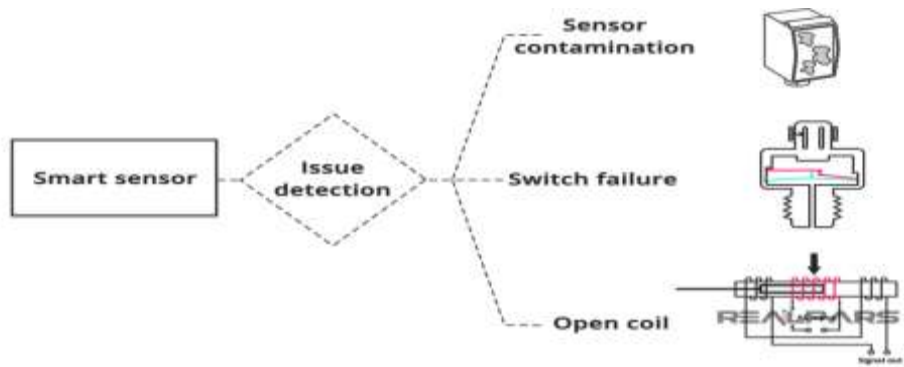

Fig-8. Technology of smart sensors

Numerous experts have recognized and acknowledged the significance of sensors, measurement science, and smart assessment for Industry 4.0, which has already resulted in the declaration "Industry 4.0: nothing goes without sensor systems[22]."

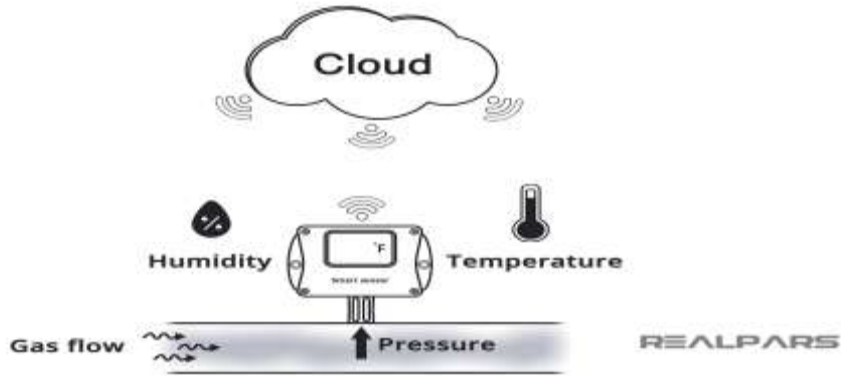

Fig-9. Application of smart sensors

Some multi-sensing smart sensors can detect pressure, temperature, humidity, gas flow, and other things. The implementation of complete systems is, however, constrained by a few problems, including time synchronization errors, data loss, and handling huge amounts of gathered data. Additionally, additional restrictions on these features correspond to the battery power[23]. Smartwatches are one example of how smart sensors are being integrated into electronic devices. In this case, sensors collect data from the user's movement, process it, and then give the user information about how many steps they have taken throughout the day as well as convert the data into calories burned[24].

## 3.4. Quantum Computing

A promising new technology called quantum computing has the potential to accelerate the development of AI. Unlike binary digital electronic computers, which are built on transistors, quantum computers carry out a new kind of computation. Data is often encoded into binary digits (bits), each of which is always in one of two specified states during common digital computing (0 or 1)[25].

There are two possible measurements of a qubit: 0 and 1. Therefore, the qubit is a two-dimensional quantum system. Each dimension is denoted by a standard basis vector. In quantum computing, we use the Dirac notation. It represents a column vector by the ket that looks like "$|\psi\rangle$":

$$|0\rangle = \begin{bmatrix} 1 \\ 0 \end{bmatrix} \text{ in Python [1, 0] and}$$

$$|1\rangle = \begin{bmatrix} 0 \\ 1 \end{bmatrix}, \text{ in Python [0, 1].}$$

The state of the qubit is represented by the superposition of both dimensions. This is the qubit state vector $|\psi\rangle$ ("psi").

$$|\psi\rangle = \alpha|0\rangle + \beta|1\rangle = \begin{bmatrix} \alpha \\ \beta \end{bmatrix}, \text{ with } \alpha^2 + \beta^2 = 1$$

α and β are the probability amplitudes of the states $|0\rangle$ and $|1\rangle$. Their squares denote the probabilities of measuring the qubit as 0 (given by α^2) or 1 (β^2) respectively. $|\psi\rangle$ must be normalized by:

$$\alpha^2 + \beta^2 = 1$$

Let's have a look at a graphical representation of the qubit state |ψ⟩ in the following figure. In this representation, both dimensions reside at the vertical axis but in opposite directions. The top and the bottom of the system correspond to the standard basis vectors |0⟩ and |1⟩, respectively.

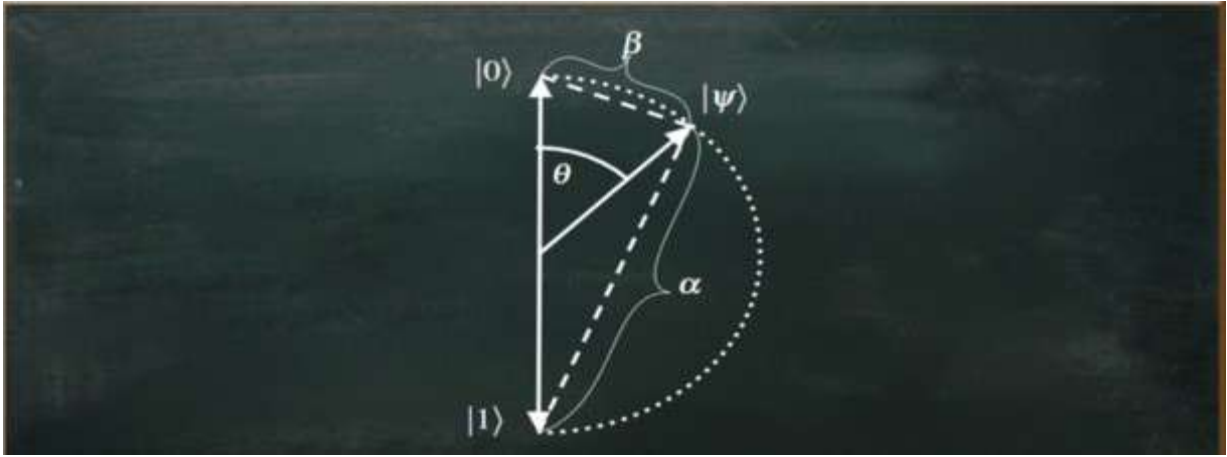

Fig-10. 2-dimensional qubit system

Contrarily, quantum computation employs quantum bits, which are capable of existing in several superpositions of states. Because of their increased speed and ability to handle vast amounts of complex data, they will soon be able to run large-scale simulations like those carried out by physicists at CERN's Hadron collider [26].

Since the amount of data we produce is growing at an exponential rate, it is projected that quantum computing will advance much more shortly. This implies that to figure out what to do with such enormous amounts of data, we will likewise want much more powerful computers[27].

The two most promising branches of AI promoting the 4IR, machine learning, and deep learning, are anticipated to be impacted to ever higher heights by quantum computing and neuromorphic devices.

International Journal of Computer Science & Information Technology (IJCSIT) Vol 14, No 4, August 2022

## 3.5. IoT(Internet of Things)

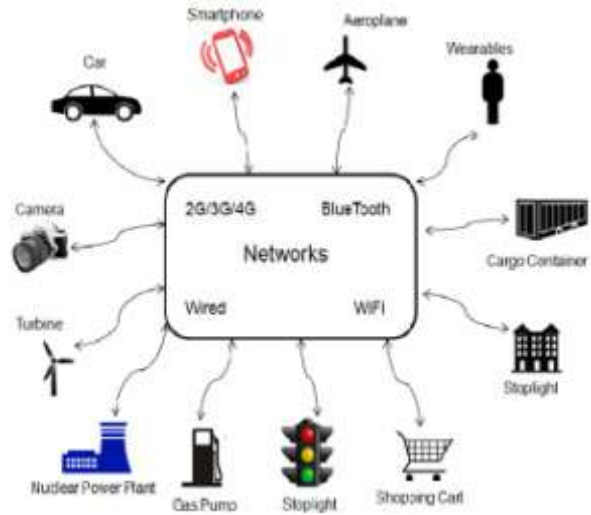

Fig-11. Internet of Things(IoT)

The term "Internet of things" (IoT) refers to physical objects (or groups of such objects) equipped with sensors, computing power, software, and other technologies that communicate with one another and exchange data through the Internet or other communications networks. The term "internet of things" has been considered a misnomer because devices only need to be individually addressable and connected to a network, not the whole Internet[28].

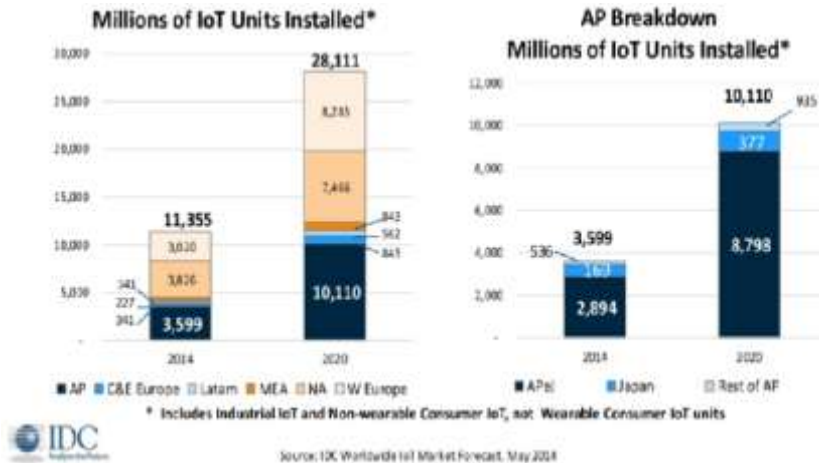

Fig-12. The Internet of Things Market Size

The fusion of numerous technologies, such as ubiquitous computing, widely available sensors, sophisticated embedded systems, and machine learning, has caused the sector to advance. The Internet of things is enabled by the traditional fields of embedded systems, wireless sensor networks, control systems, and automation (including home and building automation)[29]. IoT products are most often associated with the "smart home" in the consumer market because they support one or more common ecosystems and can be controlled by gadgets related to those ecosystems, like smart speakers and smartphones. These products include lighting fixtures, thermostats, home security systems, cameras, and other appliances. Systems for providing healthcare also leverage IoT [30].

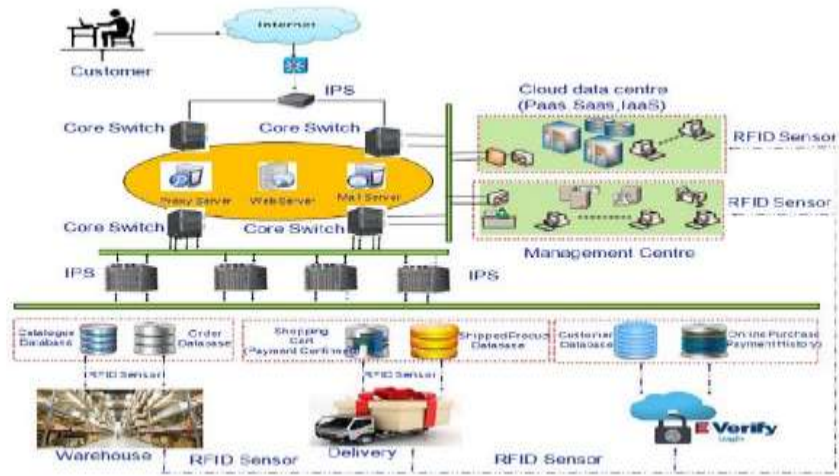

Fig-13. Reference IoT Architecture for E-commerce

Recent technological developments have had a significant impact on the e-commerce industry. It is crucial for the businesses involved to use the appropriate technology to satisfy their clients, especially when it comes to IoT, as consumers' unpredictable lifestyles transform and they quickly get fond of online purchasing. IoT-enabled devices exchange data with one another within an internet network, assisting both retail and e-commerce businesses in conducting their operations as easily and effectively as possible[31].

The way that customers buy has drastically changed with the introduction of IoT devices like smart mirrors that allow users to easily try on garments virtually and are coupled with technology to help users re-purchase products of their choosing. The creation of IoT applications has become extremely popular over time. We'll talk about how the Internet of Things (IoT) is crucial to the e-commerce sector in this blog[32].

## 3.6. AI (Artificial Intelligence)

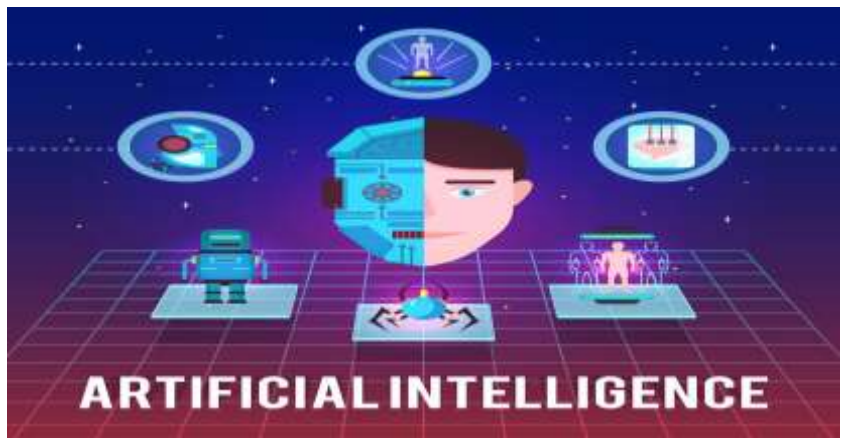

Fig-14. Artificial Intelligence

In contrast to the natural intelligence exhibited by animals, including humans, artificial intelligence (AI) is intelligence demonstrated by robots. The study of intelligent agents, or any system that understands its environment and acts in a way that maximizes its chances of succeeding, has been defined as the focus of AI research. Previously, robots that mimic and exhibit "human" cognitive abilities associated with the human mind, like "learning" and "problem-solving," were referred to as "artificial intelligence."Major AI researchers have now rejected this notion and are now describing AI in terms of rationality and acting rationally, which does not constrain how intelligence can be expressed[33].

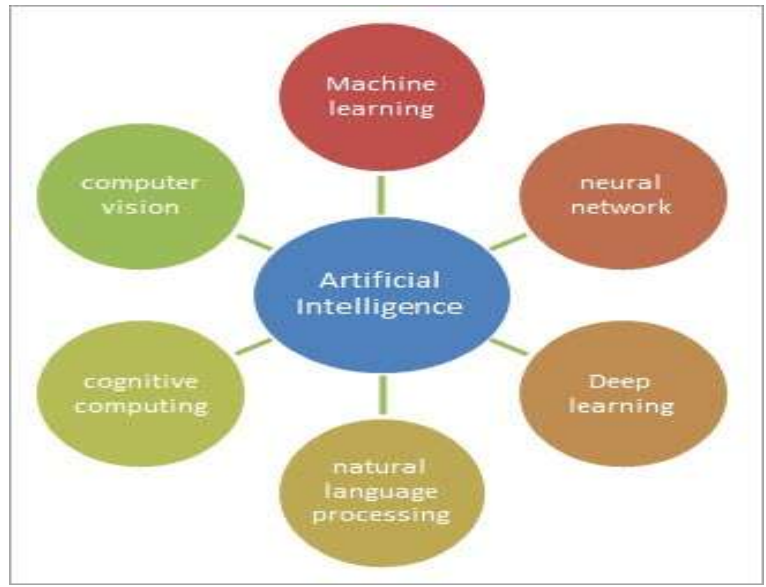

Fig-15: Subfields of artificial intelligence

A few examples of AI applications are cutting-edge web search engines like Google, recommendation systems like YouTube, Amazon, and Netflix, speech recognition software like Siri and Alexa, self-driving cars like Tesla, automated decision-making, and dominating the best strategic game systems (such as chess and Go). The AI effect is a phenomenon where actions once thought to require "intelligence" is frequently taken out of the definition of AI as machines grow more and more capable. For instance, despite being a commonplace technique, optical character recognition is typically left out of the list of items that are regarded to be AI[34].

## 4. IMPACT OF 4 IR TECHNOLOGY ON THE BUSINESS

The Fourth Industrial Revolution has four major implications for business: changing customer expectations, improving products, fostering collaborative innovation, and changing organizational structures. Customers, whether they are consumers or enterprises, are more and more at the center of the economy, which revolves around enhancing the customer experience. Additionally, it is now possible to add digital capabilities to physical goods and services to raise their value. While data and analytics are revolutionizing asset maintenance, new technologies are enhancing asset resilience and durability[35]. New forms of collaboration are necessary in a world of customer experiences, data-based services, and asset performance through analytics. This is especially true given how quickly innovation and disruption are occurring. Finally, the advent of global platforms and other new business models calls for a reevaluation of people, culture, and organizational structures.

Overall, the Fourth Industrial Revolution is driving businesses to reevaluate how they conduct business due to the inexorable transition from basic digitalization (the Third Industrial Revolution) to innovation based on a combination of technologies. The fundamental truth, however, remains the same: company executives and senior leaders must comprehend their shifting environment, question the presumptions of their operating teams, and innovate ceaselessly and ruthlessly.

## 5. IMPACT OF 4IR TECHNOLOGY ON THE GOVERNMENT

The nature of domestic and global security will be significantly altered by the Fourth Industrial Revolution, which will also have an impact on the likelihood and specifics of any conflict. Today is no exception to the rule that the history of conflict and global security is also the history of technological advancement. Modern state-on-state confrontations are becoming more "hybrid" in nature, fusing conventional military tactics with components formerly connected to non-state actors. Uncomfortably, lines are blurring between violence and nonviolence (think of cyberwarfare), combatant and noncombatant, and even war and peace [36].

Individuals and small groups will increasingly join states in being capable of wreaking havoc as this process progresses and new technology like autonomous or biological weapons become simpler to utilize. Fears will arise as a result of this new vulnerability. However, at the same time, technological advancements offer the possibility of reducing the scope or impact of violence, for instance through the creation of novel forms of defense or improved targeting.

## 6. IMPACT OF 4IR TECHNOLOGY ON THE PEOPLE

Finally, the Fourth Industrial Revolution will alter both what we do and who we are. Our feeling of privacy, our concepts of ownership, our purchasing habits, the amount of time we spend working and playing, and how we pursue careers, improve our talents, meet people, and foster relationships will all be impacted by this. Our health is already changing as a result of it, and it is creating a "quantified" self. Eventually, it might even lead to human enhancement. The possibilities are unlimited because they are only limited by our imagination[37].

Privacy is one of the biggest issues that new information technologies bring to individuals. Although we intuitively understand why it is so important, a key component of the new connectivity is the tracking and exchange of information about us. In the coming years, discussions regarding important topics like how losing control of our data will affect how we live our inner lives will get more heated. Similar to how biotechnology and artificial intelligence (AI) breakthroughs are changing what it means to be human by advancing the present limits of life expectancy, health, intellect, and capabilities, these developments will force us to reevaluate our moral and ethical standards[38].

## 7. CONCLUSION

Finally, the Fourth Industrial Revolution will alter both what we do and who we are. Our feeling of privacy, our concepts of ownership, our purchasing habits, the amount of time we spend working and playing, and how we pursue careers, improve our talents, meet people, and foster relationships will all be impacted by this. The Fourth Industrial Revolution, also known as Industry 4.0 or Industrie 4.0, is a technological revolution that is taking place in the modern developing environment where novel technologies and trends like virtual reality (VR), the Internet of Things (IoT), artificial intelligence (AI), and robotics are fundamentally changing how people live, work, and interact with one another. Every aspect of human existence is changing as a result of the Fourth Industrial Revolution, from the government to business to healthcare. By altering both human beings' virtual and actual physical worlds, it is even having an impact on their values, opportunities, relationships, and identities.

International Journal of Computer Science & Information Technology (IJCSIT) Vol 14, No 4, August 2022[21] Xaba, S. A., Fang, X., & Mthembu, S. P. (2021). The Impact of the 4IR Technologies in the Works of Emerging South African Artists. *Art and Design Review*, *9*(1), 58-73.

[22] Mtshali, T. I., & Ramaligela, S. M. (2020). Contemporary employability skills needed for learners to succeed in the civil technology field in the 4IR era. *Journal of Technical Education and Training*, *12*(3), 29-40.

[23] Ebekozien, A., Aigbedion, M., Duru, O. S. D., Udeagwu, O. H., & Aginah, I. L. (2021). Hazards of wood sawmills in Nigeria's cities: the role of fourth industrial revolution technologies. *Journal of Facilities Management*.

[24] Andreoni, A., Chang, H. J., & Labrunie, M. (2021). Natura non facit saltus: Challenges and opportunities for digital industrialisation across developing countries. *The European Journal of Development Research*, *33*(2), 330-370.

[25] Mpofu, R., & Nicolaides, A. (2019). Frankenstein and the fourth industrial revolution (4IR): ethics and human rights considerations. *African Journal of Hospitality, Tourism and Leisure*, *8*(5), 1-25.

[26] Nobre, I., & Nobre, C. A. (2019). The Amazonia third way initiative: the role of technology to unveil the potential of a novel tropical biodiversity-based economy. *Land Use—Assessing the Past, Envisioning the Future*, 1-32.

[27] Kruger, S., & Steyn, A. A. (2020). Enhancing technology transfer through entrepreneurial development: practices from innovation spaces. *The Journal of Technology Transfer*, *45*(6), 1655-1689.

[28] Lee, S. W., Jo, J., & Kim, S. (2021). Leveraging the 4th Industrial Revolution technology for sustainable development of the Northern Sea Route (NSR)—The case study of autonomous vessel. *Sustainability*, *13*(15), 8211.

[29] Herweijer, C., Combes, B., Johnson, L., McCargow, R., Bhardwaj, S., Jackson, B., &Ramchandani, P. (2018). *Enabling a sustainable Fourth Industrial Revolution: How G20 countries can create the conditions for emerging technologies to benefit people and the planet* (No. 2018-32). Economics Discussion Papers.

[30] Kim, H. S., & Hwang, W. S. (2022). Absorption trajectories of 4IR technologies: evidence from Korea. *Technology Analysis & Strategic Management*, 1-17.

[31] Bigerna, S., Micheli, S., & Polinori, P. (2021). New generation acceptability towards durability and repairability of products: Circular economy in the era of the 4th industrial revolution. *Technological Forecasting and Social Change*, *165*, 120558.

[32] Shilenge, M., & Telukdarie, A. (2021). 4IR integration of information technology best practice framework in operational technology. *Journal of Industrial Engineering and Management*, *14*(3), 457-476.

[33] Jin, B. E., & Shin, D. C. (2021). The power of 4th industrial revolution in the fashion industry: what, why, and how has the industry changed?. *Fashion and Textiles*, *8*(1), 1-25.

[34] Bhattacharya, S., & Chatterjee, A. (2021). Digital project driven supply chains: a new paradigm. *Supply Chain Management: An International Journal*.

[35] Olaitan, O. O., Issah, M., & Wayi, N. (2021). A framework to test South Africa's readiness for the fourth industrial revolution. *South African Journal of Information Management*, *23*(1), 1-10.

[36] Tugizimana, F., Engel, J., Salek, R., Dubery, I., Piater, L., & Burgess, K. (2020). The disruptive 4IR in the life sciences: Metabolomics. In *The Disruptive Fourth Industrial Revolution* (pp. 227-256). Springer, Cham.

[37] Tella, A., Amuda, H. O., & Ajani, Y. A. (2022). Relevance of blockchain technology and the management of libraries and archives in the 4IR. *Digital Library Perspectives*.

[38] Evron, Y. (2021). 4IR technologies in the Israel Defence Forces: blurring traditional boundaries. *Journal of Strategic Studies*, *44*(4), 572-593.

[39] Berlanstein, Lenard R., ed. *The industrial revolution and work in nineteenth century Europe*. Routledge, 2003.

[40] Forrester, Rochelle. "History of electricity." *Available at SSRN 2876929* (2016).

[41] Janicke, Martin, and Klaus Jacob. "A third industrial revolution." *Long-term governance for social-ecological change* (2013): 47-71.
66

**AUTHORS**

**Bandar alsulaimani** is currently a Ph.D. researcher from the Faculty of Electrical Engineering, King Fahad University of Petroleum and Minerals (KFUPM), Saudi Arabia. Bandar is a Saudi computer engineer. He studied computer networks at King Fahad University of petroleum and minerals (KFUPM) and graduated with a master's degree as well from the same University. His research interests include computer networks, network security, network system,4IR technology, and wireless system.

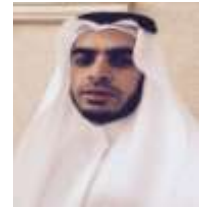

**Amanul islam** is currently a Research Assistant from the Faculty of Computer Science and Information Technology, University of Malaya, Malaysia. He received a BS degree in Information and Communication Technology from The Millennium University, Bangladesh. He received his master's degree in Computer Science By Research from the Faculty of Computer Science and Information Technology, University of Malaya, Malaysia. His research interests include computer networks, network security, graphical authentication, and wireless system.

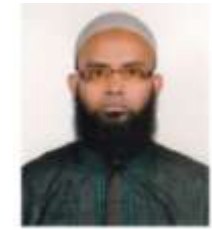